\documentclass [11pt,a4paper]{article}
\pdfoutput=1 
\usepackage{jheppub}
\usepackage{latexsym,epsfig,amssymb, amsmath,nicefrac}

\usepackage{subcaption}

\usepackage{slashed}

\usepackage{color}
\usepackage[makeroom]{cancel}
  \usepackage{latexsym}
  \usepackage{epsf}
  \usepackage{amssymb}
  \usepackage{graphicx}
  \usepackage{amsmath}
  \usepackage{amsmath,amssymb,amsthm}
  \usepackage{verbatim}
  \usepackage{hyperref}

\def\tb0{\tilde{\beta}_0}
{\def\b0{\beta_0}

\def\bi{\begin{itemize}}
\def\ei{\end{itemize}}
\def\be{\begin{equation}}
\def\ee{\end{equation}}
\newcommand{\bea}{\begin{eqnarray}}
\newcommand{\eea}{\end{eqnarray}}
 
\renewcommand{\Im}{\textrm{Im}\,}
\renewcommand{\Re}{\textrm{Re}\,}

\begin{document}

\begin{flushright}
MPP-2022-112
\end{flushright}

\title{Towards Early Dark Energy in String Theory}

\author{Evan McDonough$^1$}
\emailAdd{e.mcdonough@uwinnipeg.ca}
\affiliation{$^1$Department of Physics, University of Winnipeg, Winnipeg MB, R3B 2E9, Canada }

\author{and Marco Scalisi$^2$}
\emailAdd{mscalisi@mpp.mpg.de}
\affiliation{$^2$Max-Planck-Institut für Physik (Werner-Heisenberg-Institut),
Föhringer Ring 6, 80805, München, Germany}

\abstract{Early Dark Energy (EDE) is a prominent model to resolve the Hubble tension, which employs a dynamical axion with a periodic potential.  In this work, we take first steps towards the embedding of this model into stable compactifications of string theory. First, we provide a pedagogical review of the EDE scenario and its main challenges. Second, we construct a simple supergravity toy model using only minimal ingredients. Already at this level, we can understand the origin of the harmonics of the EDE scalar potential in terms of a delicate balance of the leading terms from separate non-perturbative effects. Third and final, we embed the model into a KKLT-type compactification, with the EDE scalar field realized by a two-form axion. We find that a successful embedding, with all moduli stabilized, requires restrictive assumptions both on the Pfaffians and on the exponents of the non-perturbative terms responsible for the EDE dynamics. We point out that such non-generic conditions reflect well known challenges of the EDE model and further investigation might guide us towards a conclusive resolution.
}

\maketitle


  \section{Introduction: Challenges for Early Dark Energy}

With three Nobel prizes to its credit, the Cosmic Microwave Background (CMB) is one of the crown jewels of modern cosmology. By measuring angles in the CMB snapshot of the early universe, and combining this with the relative abundances of matter, radiation and dark energy, one may infer the present expansion rate of the universe $H_0$. The Planck 2018 experiment reports \cite{Aghanim:2018eyx}
 \begin{equation}
H_0 = 67.37 \pm 0.54 \, {\rm km/s/Mpc}\,.
\end{equation}
 An orthogonal approach, with its own Nobel prize, is to build a {\it cosmic distance ladder}, namely the SH0ES collaboration \cite{Riess:2019cxk,Riess:2020fzl} measurement using Type Ia supernovae (SNIa) \cite{Riess:2021jrx}, which gives 
 \begin{equation}
H_0 = 73.04 \pm 1.04 \, {\rm km/s/Mpc}\,.
\end{equation} 
The SH0ES measurement  and Planck CMB inference are in $5\sigma$  disagreement \cite{Riess:2021jrx}, meeting the threshold statistical significance usually associated with a discovery in particle physics. The so-called {\it Hubble tension} persists when other cosmological data sets are considered, e.g. large-scale structure (LSS) \cite{2016ApJ...830..148C,Aubourg:2014yra,Cuceu:2019for,Schoneberg:2019wmt,Abbott:2017smn,Philcox:2020vvt}, and from other probes \cite{Verde:2019ivm}, but is most acute when Planck vs. SH0ES are considered. While experimentalists and observational cosmologists refine detector sensitivities and search for an explanation in systematic errors, the sheer magnitude of the disagreement behooves theorists to contribute on their frontier. To this end, the main challenge is to understand whether the discrepancy is an artifact of the $\Lambda$CDM  model.

A prominent proposal for an alternative cosmological model is {\it Early Dark Energy} (EDE)  \cite{Poulin:2018cxd}. In this scenario, the expansion rate $H(t)$ is increased shortly before the CMB formed, which in turn raises the inferred value $H_0$ from CMB data. We explain this in detail in Sec.~\ref{sec:EDE}. This model is described by a scalar field $\phi$ with potential,
 \begin{equation}
 V_{\rm EDE} = \Lambda^4 _{\rm EDE}\left( 1 - \cos \frac{\phi}{f}\right)^n ,
 \end{equation} 
 with power $n \geq 2$, energy scale $\Lambda_{\rm EDE} \sim {\rm eV}$  and decay constant $f \sim 0.2\ M_{pl}$ ($M_{pl}=2.435 \times 10^{18}\, {\rm GeV}$ is the reduced Planck mass).
The cosmological dynamics of $\phi$ are very similar to a conventional ultra-light axion \cite{Marsh:2015xka}: at early times the field is frozen in place by Hubble friction, and effectively behaves as dark energy. At a later time, the field rolls down its potential and begins to oscillate in the minimum. The exponent $n$ controls the shape of the potential about the minimum, and in turn the rate at which the energy density in $\phi$ is redshifted away; $n\geq 2$ in order for EDE to address the Hubble tension, else it is completely degenerate with adding an ultra-light axion dark matter ($n=1$), which is already strongly constrained \cite{Hlozek:2014lca,Lague:2021frh}. The best-fit model is given by $n=3$ \cite{Poulin:2018cxd}, which has in fact been the focus of much of the literature (see e.g. \cite{Smith:2019ihp,Hill:2021yec,Reeves:2022aoi,Herold:2021ksg,Fondi:2022tfp,Klypin:2020tud}) and will also be the target  of the present work. We note there have recently been developed many EDE-like models (e.g., \cite{Agrawal:2019lmo,Alexander:2019rsc,Lin:2019qug,Sakstein:2019fmf,Niedermann:2019olb,Niedermann:2020dwg,Kaloper:2019lpl,Berghaus:2019cls,Alexander:2022own}), which we will not focus on here. \\

Despite its simplicity and success in resolving the Hubble tension, the EDE model comes also with a number of hurdles. The three main challenges\footnote{An additional puzzle, discussed in detail in \cite{McDonough:2021pdg}, is the near-Planckian decay constant, and consistency with the Swampland Distance Conjecture \cite{Ooguri:2006in}. } of this scenario are:
 \begin{enumerate} 
 
\item[]{\bf Challenge I: Energy scale.} The timing of the epoch when the EDE is cosmologically relevant, namely around matter-radiation equality with a redshift of $z_{\rm eq} \sim 3000$,  fixes the energy scale of the EDE as $\Lambda_{\rm EDE} \sim {\rm eV}$. The challenge of realizing such a small energy density is nearly as severe as the challenge of obtaining a cosmological constant with energy density $\rho_{\rm \Lambda} ^{1/4} \sim {\rm meV}$. While there are well known constructions of the cosmological constant in string theory \cite{Kachru:2003aw,Burgess:2003ic,Westphal:2006tn}  (see also \cite{Kallosh:2019axr,Kallosh:2018psh,Kallosh:2018wme} for some further analysis on the 4D effective models and their relation to Swampland conjectures), there are no constructions of EDE in string theory, and no demonstration that EDE and the cosmological constant can coexist in one self-consistent UV framework.

 \item[] {\bf Challenge II: Origin of the potential.} The conventional origin of the periodic potential of an axion is the leading term in an instanton expansion. In this context,  the EDE potential with $n=3$ requires the first three harmonics to be roughly equal, and the higher-order harmonics to vanish. This implies a very delicate balance of terms  in order to have a controlled instanton expansion. This issue was recently pointed out also in \cite{Rudelius:2022gyu}.

 \item[ ]{\bf Challenge III: Large-scale structure data.}  The shifts in $\Lambda$CDM parameters necessary to maintain the fit to CMB data in EDE bring the model into tension with large-scale structure data \cite{Hill:2020osr,Ivanov:2020ril,DAmico:2020ods}; see also \cite{Jedamzik:2020zmd} and \cite{Lin:2021sfs}. The EDE model thus alleviates one tension at the cost of exacerbating an other. (Though see  \cite{Simon:2022adh,Murgia:2020ryi,Smith:2020rxx} for an alternate viewpoint). This issue will be explained in detail in Sec.~\ref{sec:EDE}.
 \end{enumerate}
In this work, we seek to elucidate these challenges in the context of a UV completion in string theory. We seek to understand the origin of the cosine-cubed potential with the appropriate energy scale, in order to contextualize and gauge the severity of the fine-tunings of the EDE model. We also seek to understand  how these same model constructions might play a role in ameliorating the tension with large-scale structure. To this end, we provide simple realizations where the $n=3$ potential can be realized in a controlled expansion, namely via a coincidence of energy scales between separate non-perturbative terms.

 The structure of this paper is as follows: We first provide a pedagogical review of the EDE model and its observational constraints. We then develop a toy model in supergravity wherein the Early Dark Energy is realized via non-perturbative terms in the superpotential, and which can simultaneously accommodate the current cosmological constant. We then develop a string theory model utilizing gaugino condensation on D5 branes to generate a periodic potential for a $C_2$ axion, similar to the natural inflation construction in \cite{Ben-Dayan:2014lca}. We consider a KKLT-like compactification, and explicitly demonstrate that volume stabilization at a positive value of the scalar potential can be achieved simultaneously with the EDE phase. The fine-tuning of the EDE energy scale is inherited by the energy scale of the non-perturbative terms in the superpotential. We close with a discussion of directions for future work, both for EDE model building and string theory constructions.

\section{Review of Early Dark Energy}

\label{sec:EDE}

\subsection{The EDE model}

Cosmic Microwave Background data is a snapshot in time of acoustic oscillations in the primordial photon-baryon plasma, which in turn correspond to the oscillations of primordial perturbations after horizon re-entry.  These oscillations travel a finite distance from the onset of radiation domination until last scattering, given by the sound horizon $r_s$ as
\begin{equation}
\label{eq:rs}
   r_s(z_*) = \int _{z_*} ^{z_{\rm re}} \frac{{\rm d} z}{H(z)} c_s(z) \,,
\end{equation}
where $z_* \approx 1100$ is the redshift of last scattering, $z_{\rm re}$ is the redshift of reheating after cosmic inflation, and $c_s(z)$ is the sound-speed of fluctuations in the primordial plasma. The value of $r_s$ is predominantly set by cosmic evolution in the decade of redshift leading up to last scattering \cite{Knox:2019rjx}. The evolution of $H(z)$ is determined by the energy density in each component of the universe as,
\begin{equation}
H(z) = H_0 \sqrt{ \Omega_m (1+z)^3 + \Omega_{\rm rad}(1+z)^4 + \Omega_{\Lambda} + ...} \,\,,
\end{equation}
where $...$ corresponds to additional components not contained in the $\Lambda$CDM model. The sound-speed $c_s$ depends on the relative densities of baryonic matter and radiation.

\begin{figure}[h!]
\centering
\includegraphics[width=0.8\textwidth]{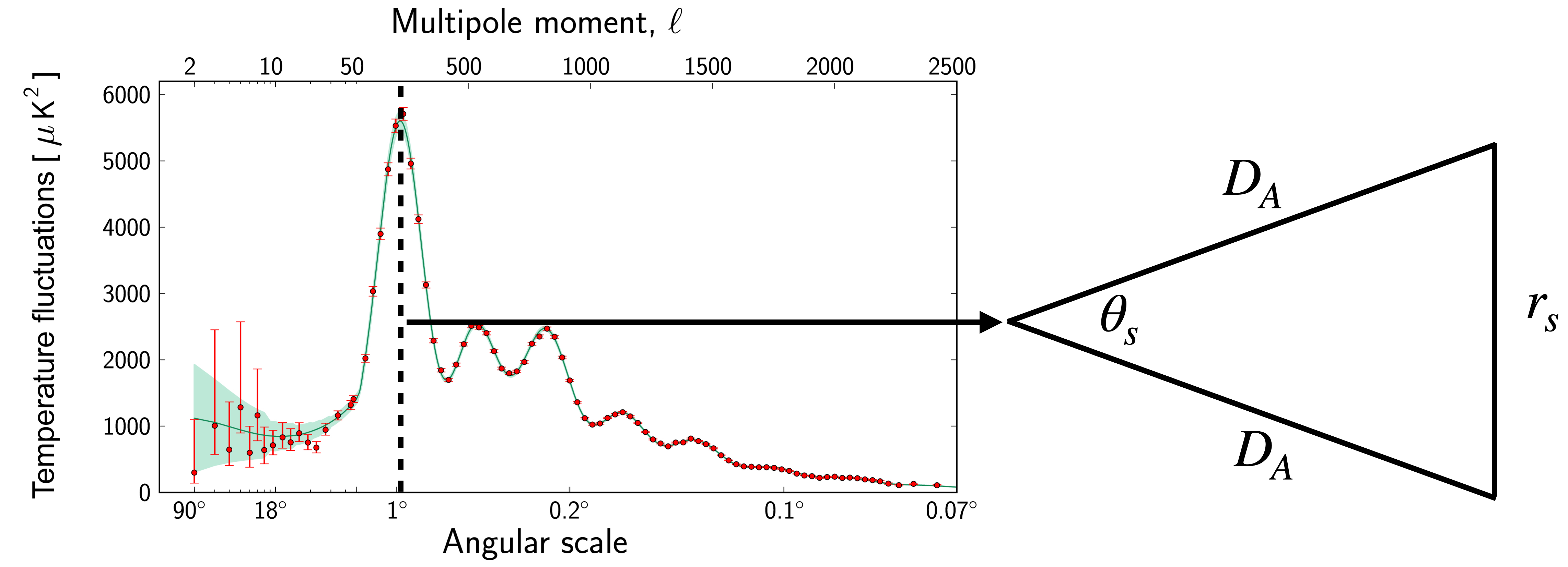}
\caption{CMB data and the sound horizon. Left Panel: Planck 2018 CMB data, namely the variance of temperature fluctuations as a function of the angle on the sky (mulitipole number $\ell$). Right panel: the mapping to the sound horizon $\theta_s$. Image adapted from \cite{HutererReview}.}
\label{fig:CMB-explainer}
\end{figure}

CMB data does not constrain $r_s$ directly, but instead constrains only the corresponding {\it angle}, namely the comoving angular diameter distance,
\begin{equation}
\label{eq:thetas}
    \theta_s = \frac{r_s (z_*)}{D_A(z_*)}\,,
\end{equation}
where $D_A$ is the angular diameter distance to last scattering, roughly speaking the distance from an observer to the surface of last scattering, given by the integral
\begin{equation}
D_A (z_*) = \int _0 ^{z_*} {\rm d}z \frac{1}{H(z)}\,,
\end{equation}
which is sensitive to $H_0$, namely the present expansion rate of the universe. One may appreciate that $\theta_s$ is the angle in an isosceles triangle with sides of length $r_{s}$ and $D_A$. This is illustrated in Fig.~\ref{fig:CMB-explainer}, adapted from \cite{HutererReview}.

 The angular extent of the sound horizon $\theta_s$ is exquisitely measured by CMB data: the \emph{Planck} 2018 analysis finds a 0.03\% measurement, $100 \theta_s = 1.0411 \pm 0.0003$ \cite{Aghanim:2018eyx}.   See Fig.~\ref{fig:CMB-explainer}. Qualitatively, $\theta_s$ can be read off from the position (multipole moment) of the first peak of the CMB temperature anisotropy power spectrum. Meanwhile, one may infer from the above expressions that raising $H_0$, e.g., to match the SH0ES measurement, reduces $D_A$. To conserve $\theta_s$, and thereby maintain the fit to CMB data, it is necessary to accordingly reduce $r_s$, e.g., by raising $H(z)$, in particular in the range $z\gtrsim z_*$, to which $r_s$ is most sensitive. 
 
 This can be accomplished through an additional contribution to the energy density of the early universe, e.g., a new dark energy like component. Whatever this new component, it must decay away very quickly after accomplishing its task,  and leave structure formation unchanged. Moreover, there are two additional angles that are well constrained by CMB data, corresponding to the Silk damping scale $r_d$ and the Hubble horizon at matter-radiation equality $r_{\rm eq}$. These are sensitive to the precise redshift dependence of any new ingredients in the $\Lambda$CDM model, and hence the redshifting of the new component.

\begin{figure}[h!]
\centering
\includegraphics[scale=0.19]{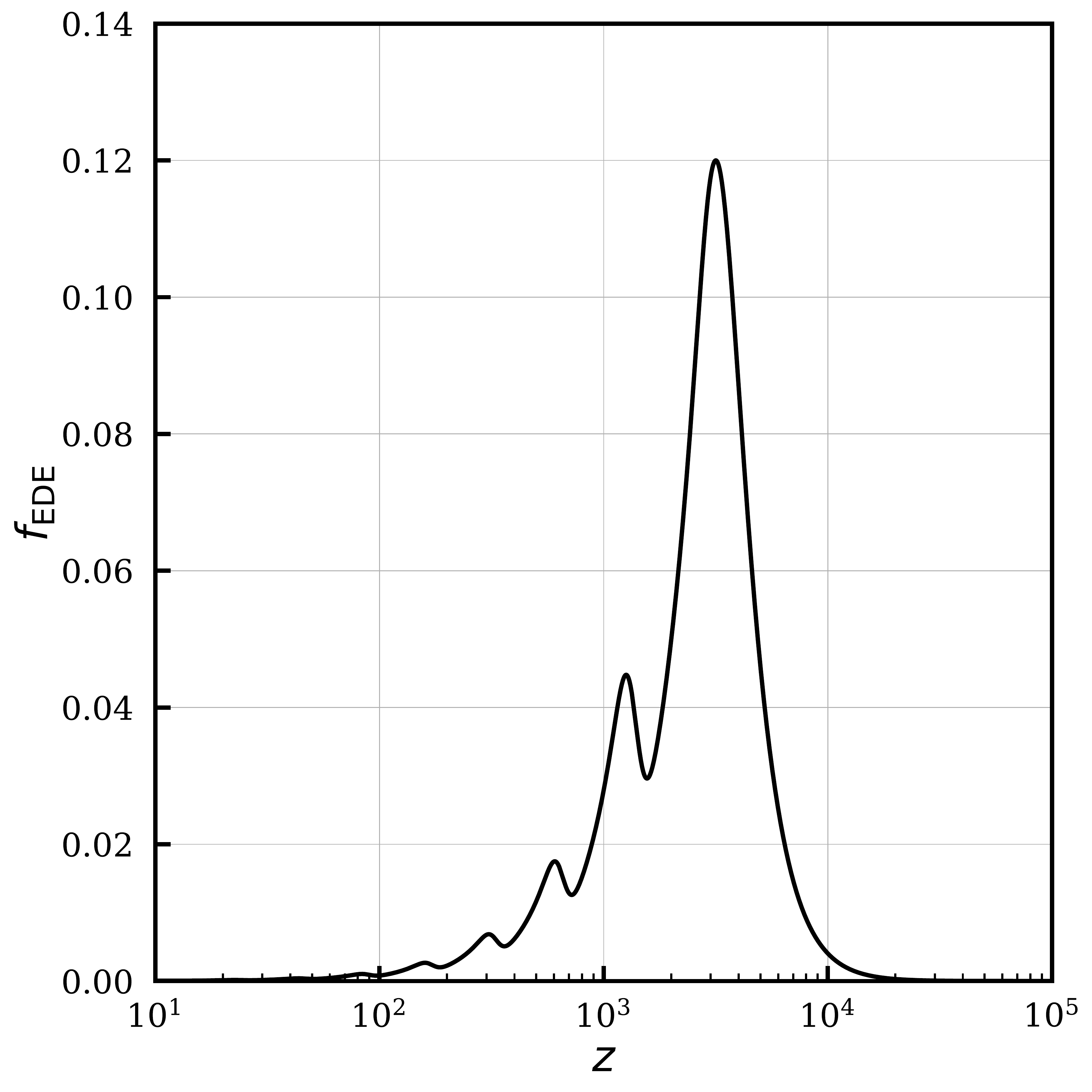}
\caption{Evolution of $f_{\rm EDE}$ in a fiducial model example with $f_{\rm EDE} = 0.12 $ and $z_c =10^{3.5}$.}
\label{fig:fEDE}
\end{figure}

A simple model realization of this idea is Early Dark Energy \cite{Poulin:2018cxd}. This scenario posits the existence of an axion-like particle with a generalized periodic potential,
\be
\label{eq:EDE_V}
V(\phi) = V_0 \left( 1- \cos \frac{\phi}{f}\right)^3 \, , \qquad V_0 \equiv m^2 f^2\,,
\ee
of the form first proposed in \cite{Kamionkowski:2014zda}.  The cosmological evolution of $\phi$ is dictated by the Klein-Gordon equation,
\begin{equation}
\ddot{\phi} +3 H \dot{\phi} + V_{,\phi} =0 \,.
\end{equation}
At early times, the field $\phi$ is frozen in place by Hubble friction, and acts as dark energy. The field rolls down its potential once $V '' \sim 3 H^2$. The field subsequently oscillates about the minimum of the potential, and the early dark energy redshifts away. This behaviour can be parametrized in terms of the fraction of the energy density of the universe,
\begin{equation}
    f_{\rm EDE}(z)  \equiv \frac{\rho_{\rm EDE}(z)}{ \rho_{\rm tot}(z)}\, ,
\end{equation}
contributed by the EDE component. This has a maximum at a critical redshift $z_c$, and it is conventional to denote $f_{\rm EDE} (z_c)$ as simply $f_{\rm EDE}$. An example of this is shown in Fig.~\ref{fig:fEDE}, where $f_{\rm EDE}= 0.12$ and $\log_{10}(z_c)=3.5$.

The exponent in Eq.~\eqref{eq:EDE_V}, in this case $3$, controls the shape of the potential about the minimum and hence the decay of the early dark energy. The requirement of this exponent is discussed in \cite{Poulin:2018cxd}.  If the exponent were $1$, as in a conventional axion, the EDE decay would redshift as matter. For exponent $2$, like radiation, and in the limit of $\infty$, like kinetic energy. The best-fit model in the fit to data has exponent $3$ \cite{Poulin:2018cxd}, and we take this to be our baseline EDE model.

The parameters required to address the Hubble tension follow from simple considerations.  In order for the field to decay around matter-radiation equality $z_{\rm eq} \approx 3000$, i.e., shortly before $z_*$, we require $V '' (\phi_i) \sim 3 H_{\rm eq}^2$, which, assuming an initial field value $\phi_i = \mathcal{O}(f)$ by standard arguments (see, e.g., \cite{Marsh:2015xka}), corresponds to, 
\begin{equation}
    \frac{V_0}{f^2} \sim 3 H_{eq}^2 \sim 3 H_0^2 z_{eq}^{3} \sim (10^{-27} {\rm eV})^2 \,.
\end{equation}
Meanwhile, the Hubble tension is roughly $10\%$, implying a shift in $H(z_c) $ of ${\cal O}(10\%)$ in order to leave $\theta_s$ unchanged. This sets the normalization of the EDE potential as,
\begin{equation}
    V_0 \sim 0.1\times 3 H_{\rm eq}^2 M_{pl}^2 \sim ({\rm eV})^4 \, .
\end{equation}
Combining the above, we determine the decay constant as,
\begin{equation}
    f \sim M_{pl} \,.
\end{equation}
Thus the EDE scenario requires a Planckian decay constant and an energy scale \mbox{$V_0 ^{1/4}\sim $ eV}. We note that these numbers are rough estimates -- restoring factors of 2 and 3 leads to $f \sim 0.3\ M_{pl}$ -- and the precise value depends on $\phi_i$, $z_c$ and the precise value of $f_{\rm EDE}$. In practice the best-fit models have $f\sim 0.2\ M_{\rm pl}$ \cite{McDonough:2021pdg}. Thus a super-Planckian decay constant is not a requirement of the model {\it per se}, though the benchmark values of $f$ are nonetheless quite near the Planck scale. The interplay of EDE and the Swampland conjectures has been explored in \cite{McDonough:2021pdg} and \cite{Rudelius:2022gyu}.

\subsection{Resolution of the Planck-SH0ES $H_0$ tension}

\begin{figure}[h!]
    \centering
    \includegraphics[ trim={400 0 2 0}, clip, scale=0.084]{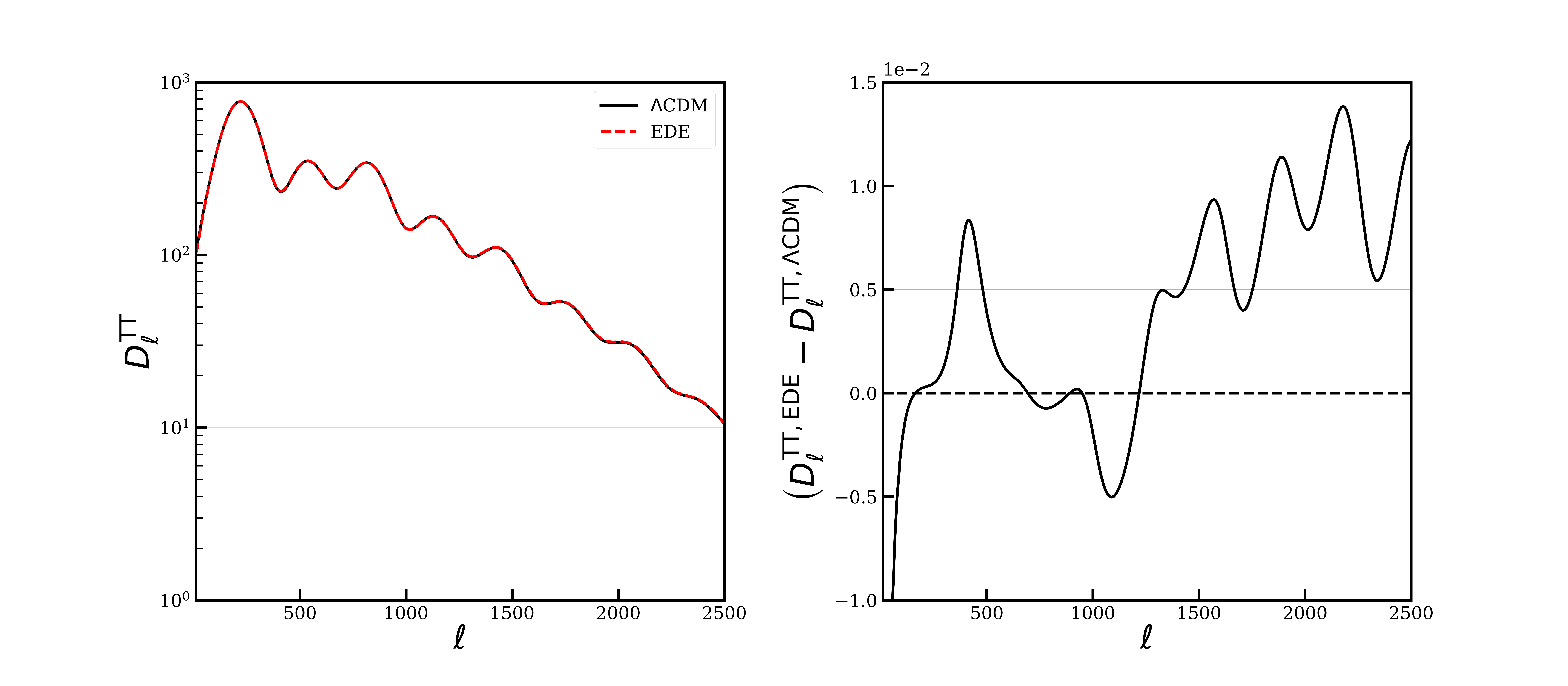}
     \caption{
    CMB temperature anisotropy power spectra (left panel) and residuals (right panel) for $\Lambda $CDM (black, solid) and EDE (red, dashed) models, with $H_0 = 68.16$ km/s/Mpc and $H_0 = 72.52$ km/s/Mpc, respectively;  parameters are the given in Table I in \cite{McDonough:2021pdg}. The curves are nearly indistinguishable in the left panel. }
    \label{fig:CMB_TT}
\end{figure}

The EDE model delivers on its promise of significantly raising $H_0$ while not degrading the fit to CMB data. This is demonstrated in Fig.~\ref{fig:CMB_TT}, where we show the angular power spectrum of CMB temperature fluctuations in EDE and $\Lambda$CDM, in the best-fit models presented in Table I of \cite{McDonough:2021pdg}, along with the fractional difference between the two, which is at the percent level.  This image is generated by the {\tt CLASS-EDE} code\footnote{Publicly available at \url{https://github.com/mwt5345/class_ede} .} \cite{Hill:2020osr}. By bringing CMB into consistency with SH0ES $H_0$ measurement, the EDE scenario resolves the tension between the CMB inference of $H_0$ and SH0ES cosmic distance ladder measurement. However, as shown in \cite{Hill:2020osr,Ivanov:2020ril,DAmico:2020ods}, and as we will review below, it does not yet provide a concordance model of cosmology.

\begin{figure}[h!]
\centering
\includegraphics[trim={400 0 2 0}, clip, scale=0.083]{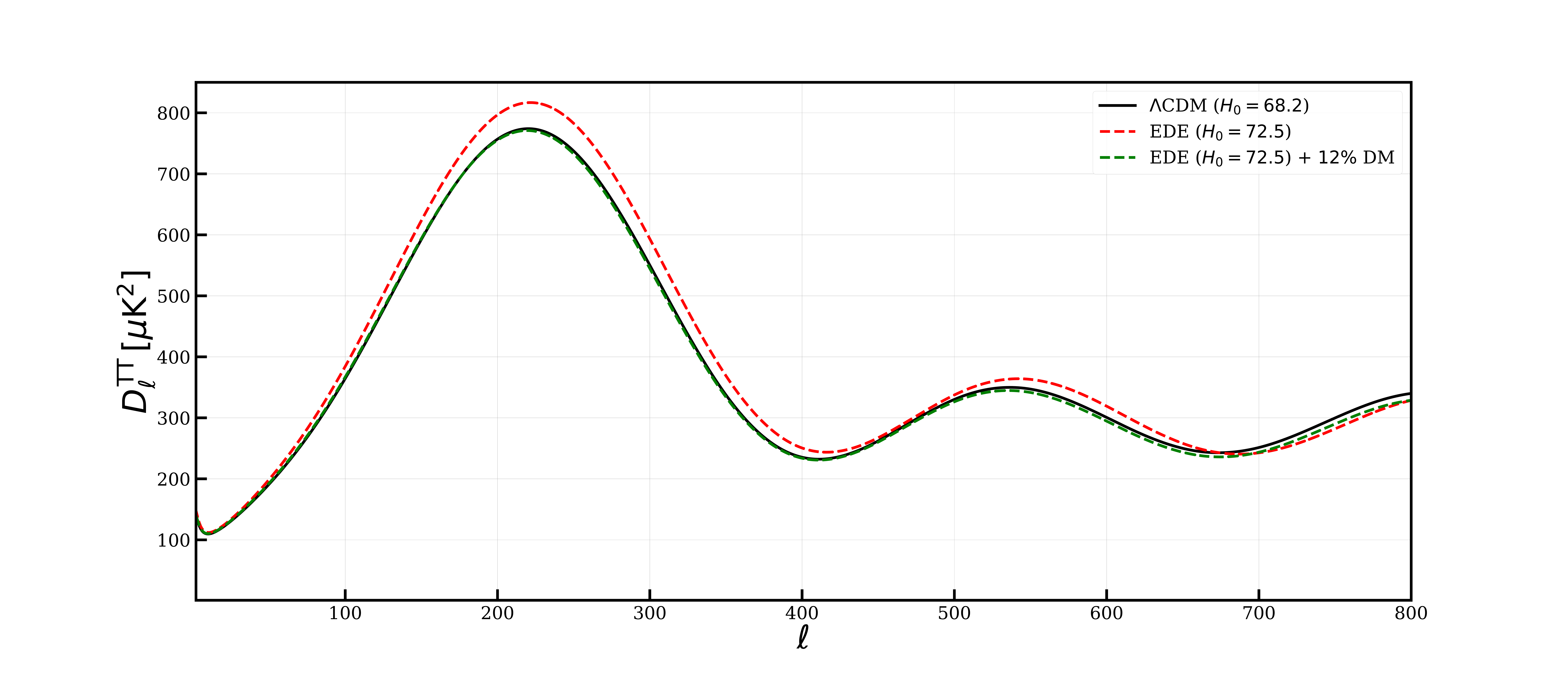}
\caption{EDE, Dark Matter, and the first peak of the CMB.  We show a fiducial $\Lambda$CDM cosmology with $H_0=68.16 $ km/s/Mpc (black), the same cosmology with $H_0$ raised to $72.52$ and with early dark energy $f_{\rm EDE}=14\%$ (red, dashed), and the same with an additional $12\%$ dark matter ($\Omega_c h^2$).  The additional dark matter is necessary to compensate the effect of EDE on the first peak. The best-fit model has additional shifts in $\Lambda$CDM parameters, e.g., $n_s$, discussed in \cite{Hill:2020osr}.}
\label{fig:CMB}
\end{figure}

EDE can provide an excellent fit to CMB data, as illustrated in Fig.~\ref{fig:CMB_TT}. However the precise mechanism for this is slightly more subtle than that presented thus far: there are compensating shifts in $\Lambda$CDM parameters, notably the amount of dark matter, as demonstrated in Fig.~\ref{fig:CMB}, and discussed in a more general context in \cite{Jedamzik:2020zmd}. As explained in  \cite{Hill:2020osr}, the extra dark energy component leads to an overall enhancement of the first peak of the CMB temperature anisotropy power spectrum. The height of the first peak is itself a measure of the amount of dark matter: dark matter acts to anchor down gravitational wells, increasing the relative redshifting of photons, and thereby dragging down the height of the first peak, while the early dark energy component has an opposite effect.

\begin{figure}[h!]
\centering
\includegraphics[width=0.9\textwidth]{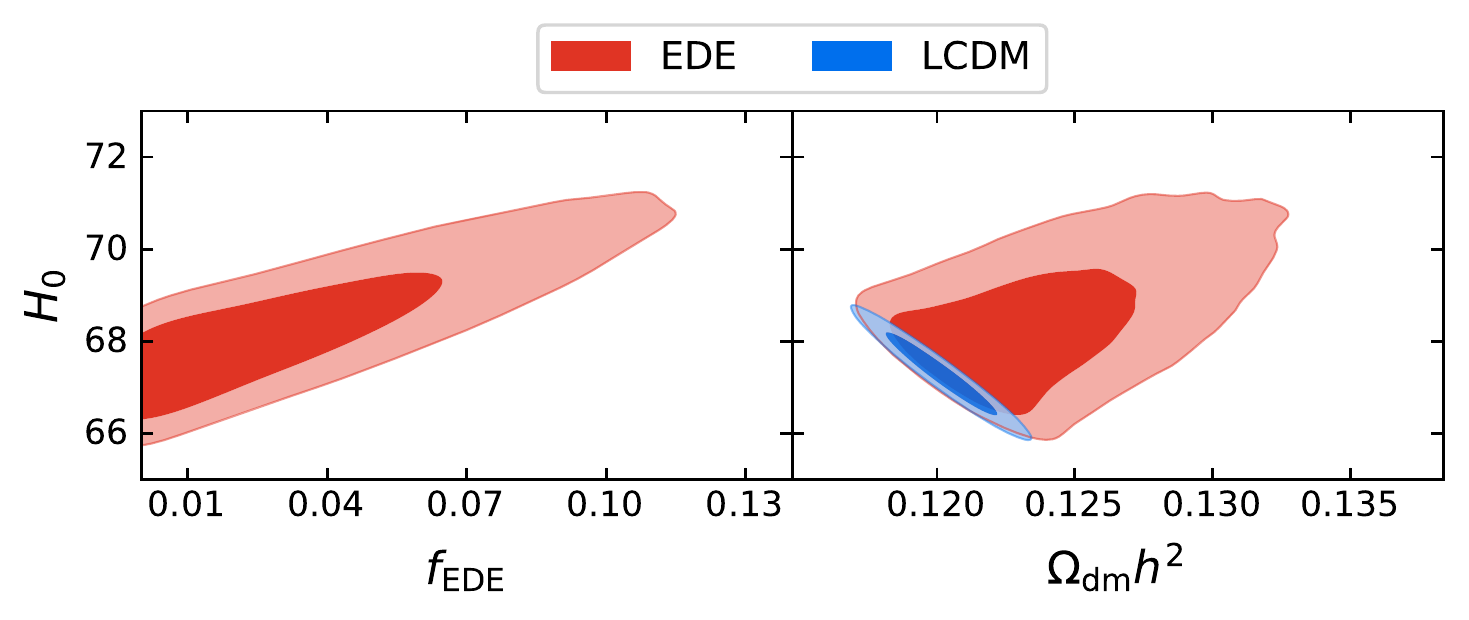}
\caption{EDE and $\Lambda$CDM posterior distributions in the fit to Planck 2018 primary CMB anisotropies. As can be anticipated from Fig~\ref{fig:CMB}, EDE raises $H_0$ by exploiting a degeneracy in $H_0$ and $f_{\rm EDE}$ (left panel), however, this brings with it additional dark matter (right panel). The $H_0-\Omega_{\rm dm} h^2$ degeneracy direction is orthogonal to that in $\Lambda$CDM; the large $H_0$ region of EDE parameter space has ${\cal O}(10\%)$ more dark matter than the range preferred by $\Lambda$CDM ($H_0 \sim 67$).}
\label{fig:data}
\end{figure}

To maintain a good fit to the height of this peak in a cosmology with ${\cal O}(10\%)$ early dark energy, one needs ${\cal O}(10\%)$ additional dark matter.  The compensation of early dark energy by dark matter is illustrated in Fig.~\ref{fig:CMB}, where the CMB temperature power spectrum is shown in a fiducial $\Lambda$CDM cosmology, in this same cosmology but with $H_0$ raised by early dark energy with $f_{\rm EDE}=14\%$, and in that same model but with dark matter raised by $12\%$. One may appreciate that the additional dark matter does an excellent job of compensating for the EDE on the first peak of the CMB.

This is borne out in the fit to the data, in the form of parameter degeneracies, namely, parameter variations that leave the fit to the data nearly unchanged, that manifest as correlations in posterior probability distributions. The analysis of Planck data in EDE was first performed in \cite{Hill:2020osr}; prior analyses had taken combined data sets, e.g., Planck in combination with the SH0ES measurement. The work \cite{Hill:2020osr}, and subsequently \cite{Ivanov:2020ril,DAmico:2020ods}, performed the first joint analysis of large-scale structure data and Planck data without SH0ES. These analyses revealed that the $H_0$-$f_{\rm EDE}$ degeneracy which allows EDE to resolve the Hubble tension is also an $H_0$-$\Omega_{\rm dm} h^2$ degeneracy, where $\Omega_{\rm dm}h^2$ is the density of dark matter.

This is clearly shown in posterior distributions in the fit to Planck, which are shown in Fig.~\ref{fig:data}. The dark and light shaded regions are $68\%$ and $95\%$ confidence regions, and red and blue denote EDE and $\Lambda$CDM respectively.  These results are publicly available\footnote{Chains publicly available at \url{https://users.flatironinstitute.org/~chill/H21_data/} .}, and can be plotted using \texttt{GetDist} \cite{GetDist}. One may appreciate from Fig.~\ref{fig:data} that the $H_0$-$f_{\rm EDE}$ contour is tilted upwards and to the right: $H_0$ can be raised without degrading the fit to Planck data, provided that $f_{\rm EDE}$ is raised with it. The $H_0$-$\Omega_{\rm dm}h^2$ contour in EDE is also tilted upward and to the right, indicating that the Planck data requires a commensurate increase in the amount of dark matter. 

This requisite additional dark matter drives tight constraints on the EDE scenario from large-scale structure data \cite{Hill:2020osr,Ivanov:2020ril,DAmico:2020ods}; see \cite{Simon:2022adh,Murgia:2020ryi,Smith:2020rxx} for further discussion and counterarguments to this. This tension with large-scale structure can be resolved by modifying the physics of dark matter: e.g., by introducing dark matter early dark energy interactions \cite{McDonough:2021pdg}, by introducing a decaying dark matter component \cite{Clark:2021hlo}, or by introducing an ultralight dark matter candidate \cite{Allali:2021azp,Ye:2021iwa} to suppress the matter power spectrum on small scales. Taken in conjunction with recent preference for EDE from the Atacama Cosmology Telescope CMB data \cite{Hill:2021yec}, there remains ample motivation to study the EDE scenario and understand the interplay of CMB and LSS.

 \section{EDE supergravity toy model} \label{SEC:SUGRA}

Our first venture is to construct a model of Early Dark Energy in supergravity. Although it has been known already for some time how to generate an arbitrary  scalar potential in the context of supergravity \cite{Kallosh:2010xz,Ketov:2014hya,Scalisi:2015qga,Ferrara:2016vzg,McDonough:2016der}, here we will use a {\it minimal approach} and employ just basic ingredients, whose string theory origin has been extensively discussed in past literature. These include non-perturbative terms in the superpotential and a nilpotent superfield\footnote{The nilpotency condition $S^2=0$ eliminates the scalar degrees of freedom and leaves just the fermionic. The absence of scalars simplifies the construction of cosmological models since one needs to stabilize less directions in moduli space. Some of the first examples were provided in \cite{Antoniadis:2014oya,Ferrara:2014kva,DallAgata:2014qsj,Kallosh:2014hxa,Scalisi:2015qga}.} $S$, which models an anti-D3 brane in a supergravity effective description\cite{Kallosh:2014wsa, Bergshoeff:2015jxa, Kallosh:2015nia,Polchinski:2015bea,Garcia-Etxebarria:2015lif,Bandos:2015xnf,Dasgupta:2016prs,Vercnocke:2016fbt,Kallosh:2016aep,Bandos:2016xyu,GarciadelMoral:2017vnz,Cribiori:2019hod,Parameswaran:2020ukp}. While the non-perturbative terms will be responsible for generating the harmonics in the scalar potential, the nilpotent field $S$ will easily add a constant contribution, which can eventually be interpreted as late-time dark energy. We will find the same ingredients also in the next section, where we embed our EDE model into a concrete scenario of moduli stabilization.

The scalar potential we would like to reproduce is the one already presented in eq.~\eqref{eq:EDE_V}, which decomposes into the following harmonics
 \begin{eqnarray}
  \label{eq:VEDE}
  V  =\Lambda+V_0 \left( 1- \cos \frac{\phi}{f}\right)^3 =  \Lambda+\frac{5}{2}V_0 - \frac{15}{4} V_0 \cos \frac{\phi}{f}  + \frac{3}{2} V_0 \cos \frac{2 \phi}{f}  - \frac{1}{4} V_0 \cos \frac{3 \phi}{f} \,,
 \end{eqnarray}
where we have added a cosmological constant term $\Lambda$.

The simplest construction involves just one chiral superfield\footnote{Constructing single-superfield models with a positive scalar potential has always represented a serious challenge for model building. This is because of the negative contribution $-3|W|^2$ to the  potential in supergravity, which tends to dominate. Successful realizations were proposed in \cite{Goncharov:1983mw,Ketov:2014qha,Ketov:2014hya,Roest:2015qya,Bernardo:2016jdr}. Our model, defined by eq.~\eqref{KWEDE}, is another working example.} $G$ with K\"ahler- and super-potential given by
\be
\begin{aligned} \label{KWEDE}
K &= \frac{1}{2} \left(G + \bar{G} \right)^2\,, \\ 
W &=  W_0  + A e^{- a G} + B  e^{- b G}\,.
\end{aligned}
\ee
where we restrict to real coefficients $A$, $B$, $a$, $b$ and $W_0$. The coefficients $W_0$, $A$, and $B$ have mass dimension $3$, while $a$ and $b$ are dimensionless. The two non-perturbative terms in $W$ will be enough to generate the three oscillatory terms of eq.~\eqref{eq:VEDE}. The K\"ahler potential is a canonical one, corresponding to flat internal curvature, and shift-symmetric in $\Im G$, which we identify as the EDE scalar field. Purely imaginary exponents $a$ and $b$ would assure always truncation exactly at $\Re G=0$. However, they are not suitable to generate sinusoidal terms, rather they would contribute with exponential terms to the scalar potential (see e.g \cite{Roest:2015qya}). In the case of real $a$ and $b$, we can still stabilize the real component of $G$ very close to the origin, as we will prove in the following.

The scalar potential is calculated as 
 \begin{equation}
 V = e^{K}\left(|D_GW|^2 - 3 |W|^2\right) . 
 \end{equation}
where $D_G W \equiv \partial_G W + W \partial_G K$  (see e.g. appendix C of \cite{Kolb:2021xfn} for standard supergravity formulae).  If we assume that we can consistently truncate our equations at $\Re G= 0$, namely the real part is massive around the origin, and define the canonically normalized field as $\phi\equiv\sqrt{2}\,\Im G$, then the set given by eq.~\eqref{KWEDE} yields
\be
\begin{aligned}
 V(\phi) = & A^2(a^2-3)+ B^2(b^2-3)-3 W_0^2 \\ 
 &-6 A W_0 \cos \left( \frac{a}{\sqrt{2}} \phi \right) +2 A B (a b-3) \cos \left(\frac{a-b}{\sqrt{2}} \phi \right ) -6 B W_0 \cos \left(\frac{b}{\sqrt{2}} \phi\right)\,. \end{aligned}
\ee 
We can now fix the parameters as,
 \begin{equation}\label{par-ab}
 a = \frac{\sqrt{2} }{f} \, ,\qquad b = \frac{3 \sqrt{2}}{f}\,,
 \end{equation}
 and
 \begin{equation}\label{par-ABW}
  A =  \sqrt{V_0}  \frac{\sqrt{15}}{2 \sqrt{\frac{2 }{f^2}-1}} \, ,\qquad B =  \frac{A}{15} \, ,\qquad W_0 =\frac{A}{6}\left( \frac{2}{f^2} - 1\right) \,.
 \end{equation}
where all quantities are in Planck units.
From this, we recover exactly the desired potential eq.~\eqref{eq:VEDE} with cosmological constant
 \begin{equation}
\Lambda =  V_ 0 \left[ \frac{729 f^4 - 324 f^2  + 100}{80 f^2 (f^2-2)}\right] \,.
\end{equation}
We notice that, within this toy model, the sign of cosmological constant strictly depends on the decay constant $f$. While it is not possible to realize a Minkowski minimum, any value of $f>\sqrt{2}$ will give a dS vacuum with $\Lambda\gtrsim \mathcal{O}(10)V_0 \sim {O}(10)\ {\rm eV}^4$. This means that this simple toy model does not allow to realize a scenario with the current observed value of dark energy density $\Lambda\simeq  {\rm meV}^4$.

In order to improve on the previous model, we can add a nilpotent superfield $S$, such to have 
\be
\begin{aligned} \label{KWEDES}
K &= \frac{1}{2} \left(G + \bar{G} \right)^2 + S\bar{S}\,, \\ 
W &=  W_0  + A e^{- a G} + B  e^{- b G} +MS\,.
\end{aligned}
\ee
where the constant $M=|D_S W|$ sets the scale of supersymmetry breaking in the $S$-direction and results into a positive contribution to the scalar potential, namely
 \begin{equation}
 V = e^{K}\left(M^2 + |D_GW|^2 - 3 |W|^2\right)\, . 
 \end{equation}
Note that no mixing term arises given the diagonal form of the K\"ahler metric. If truncation along $\Re G=0$ can be still assumed (see below for a discussion on this point), then one obtains
\be
\begin{aligned}
 V(\phi) = &M^2+  A^2(a^2-3)+ B^2(b^2-3)-3 W_0^2 \\ 
 &-6 A W_0 \cos \left( \frac{a}{\sqrt{2}} \phi \right) +2 A B (a b-3) \cos \left(\frac{a-b}{\sqrt{2}} \phi \right ) -6 B W_0 \cos \left(\frac{b}{\sqrt{2}} \phi\right) \,, \end{aligned}
\ee 
in terms of $\phi\equiv\sqrt{2}\,\Im G$. We can still fix the parameters as in eq.~\eqref{par-ab} and eq.~\eqref{par-ABW} and we obtain again the desired EDE potential
\begin{equation}
 V = \Lambda + V_0 \left( 1 - \cos \frac{\phi}{f}\right)^3 \,,
 \end{equation}
 but, this time, with tunable cosmological constant
\begin{equation}
\Lambda =  M^2 - V_ 0 \left[ \frac{729 f^4 - 324 f^2  + 100}{80 f^2 (2  - f^2)}\right] \,.
 \end{equation}
By an appropriate choice of $M$ and $f$, one can also reproduce the current observed value of the CC. Notice that the value of the decay constant should be $f<\sqrt{2}$ in order to have a delicate balance between a positive- and negative-definite term. This is in agreement with observational data, as already explained in the Introduction. 

\vspace{0.5cm}
\noindent Despite its simplicity, the model has points to watch out for:
 \begin{enumerate}

 \item The saxion $\Re G$ is stabilized at $\Re G_* \ll 1$. Its mass $m_s$, for this minimal model defined by eq.~\eqref{KWEDES}, is however `ultralight', following conventional particle physics standards. For $M = {\cal O}(V_0)$, $f\sim 0.2\ M_{pl}$ and vanishing cosmological constant $\Lambda \sim 0$, one in fact obtains $m_s \sim 10^{-26}$ eV. Such a value is comparable to that considered in the Acoustic Dark Sector model \cite{Allali:2021azp} (see also \cite{Ye:2021iwa}) to relax the tension between EDE and LSS data \cite{Hill:2020osr,Ivanov:2020ril,DAmico:2020ods}, and therefore to simultaneously resolve the $H_0$ and $S_8$ parameter tensions. Nevertheless, such a very light mass of $\Re G$ can cause some backreaction on the EDE phase (i.e. during the evolution of the EDE field $\phi\equiv\Im G$, one has a displacement $\delta\Re G_*$ which  can effectively change the form of the EDE potential). This can be avoided provided that $a {\Re}G_*, b {\Re}G_* \ll1 $, such that $e^{-a {\rm Re}G_*}$ and $e^{-b {\rm Re}G_*}$ do not change significantly from $1$. This protection from backreaction is naturally realized in the string theory setup of the next section, where the exponents are $<1$. In the present case, with $a,b \gtrsim 1$, backreaction can be prevented by the addition of a new stabilizing term in the Kahler potential, which nevertheless leaves the main features of the model invariant. One can, for example, add a correction term $\delta K = {\mu^{-2}}(G + \bar{G})^4 $ with an energy scale $\mu$, or a perturbative mixing with the nilpotent field $S$ of the form $\delta K = S \bar{S}/ \left[1 + \mu^{-2} (G + \bar{G})^2 \right]$, as originally proposed in \cite{McDonough:2016der}. These options provides additional freedom in setting the mass of Re$G$.

\item The gravitino mass of this toy model is extremely low ($m_{3/2}\simeq 10^{-54}\ \rm{eV}$). This is in strong disagreement with experimental bounds due to the lack of supersymmetric particles at energy scales of order TeV, as probed at the LHC (this implies a universal lower bound $m_{3/2}\gtrsim  \rm{meV}$). It creates also some tension with recently proposed swampland conjectures about the mass of the gravitino \cite{Kolb:2021xfn,Kolb:2021nob,Cribiori:2021gbf,Castellano:2021yye}. Also, this issue will be overcome in the next section, where the mass of the gravitino will be related to the depth of the AdS vacuum, as it is standard in any KKLT-type stabilization scenario.

\item The fine-tuning eq.~\eqref{par-ABW}  on the superpotential parameters directly reflects the request of a highly tuned potential, with very little adjustment freedom, for the realization of the EDE scenario. The aim here has in fact been a precise match with eq.~\eqref{eq:VEDE}, at {\it any} point in field space. This is unlike any attempt of de Sitter construction in supergravity or string theory, where one needs to fix the value of the potential and its first two derivatives just in one point in field space (in the case of inflation models, one has also some more freedom since most constraints come from CMB observations, which refer to a very limited part of the full potential).

\end{enumerate}

 \section{EDE in string compactifications} \label{SEC:STRING}

String theory gives a physical picture for the non-perturbative superpotential: stacks of coincident branes upon which worldvolume fermions, namely the gauginos, condense at low energies. We consider a KKLT-type compactification \cite{Kachru:2003aw}, characterized by a single 4-cycle modulus $T$, and extend this scenario to include two-form axions, as studied in  \cite{Ben-Dayan:2014lca,Blumenhagen:2015kja,Cicoli:2021tzt,Holland:2020jdh,McDonough:2018xzh}. Compactifications including two-form axions have been studied extensively, see e.g.~\cite{Cicoli:2021tzt} for a recent review. This 4d EFT is given by an ${\cal N}=1$ K\"{a}hler potential, 
 \begin{equation}
 K = - 3 \log \left[ T + \bar{T}- S \bar{S} +  \gamma(G+ \bar{G})^2\right] \,,
 \end{equation}
with two-form superfield $G=\bar{s}b_2+i c_2$, with $b_2 \equiv \int B_2$ and $c_2\equiv \int C_2$ the $B_2$ and $C_2$ fluxes, respectively, and $s$ is the axiodilaton field. The field $S$ is the nilpotent superfield of an anti-D3 brane in the KKLT scenario \cite{Kallosh:2014wsa, Bergshoeff:2015jxa, Vercnocke:2016fbt,GarciadelMoral:2017vnz,Cribiori:2019hod}. The $B_2$ axion is naturally stabilized at $\approx 0$ by the Kahler potential, and can be further stabilized by D-terms \cite{Jockers:2005zy,Grimm:2011dj}. The parameter $\gamma$ is proportional to the triple intersection numbers of the Calabi-Yau manifold; for simplicity one may consider $\gamma=-1$, though we will leave it general in what follows.

 To generate the EDE potential,  we consider gaugino condensation on D5 branes, following \cite{Ben-Dayan:2014lca}. In order not to disrupt the volume stabilization, we now utilize {\it three} non-perturbative terms for the EDE field $G$: this allows the non-perturbative cross-terms and mixing with $T$ in the scalar potential to be a small perturbative correction to the KKLT scenario. We refer to these three terms as $B$, $C$, and $D$. This can arise from D5 branes wrapping homologous cycles in a multithroated compactification or gaugino condensation in a product group $SU(N_B) \times SU(N_C) \times SU(N_D)$. The D5 branes are in addition to the D7 branes that stabilize the four-cycle volume $T$, leading to a total of 4 non-perturbative terms in the superpotential. This multitude of non-perturbative terms is familiar from many existing setups, such as the large volume scenario \cite{Conlon:2005ki,Balasubramanian:2005zx}, and the non-perturbative AdS background proposed in \cite{Bernardo:2021vfw}.

Concretely, we consider a superpotential given by,
 \begin{equation}\label{Wstring}
 W = W_0 + M S + A e^{- a T}  + B e^{- b G} +  C e^{- c G} + D  e^{- d G} ,
 \end{equation}
where the first three terms are standard in the KKLT scenario, while the $B,C,D$ terms generate the EDE dynamics. The coefficients $B,C,D$ may be computed directly in the four-dimensional supersymmetric theory, see e.g., \cite{Terning:2003th},  in the IIB string compactification (see e.g. \cite{Baumann:2006th,Haack:2006cy,Kim:2022uni}), or in M-theory, see \cite{Katz:1996th,Katz:1996fh,Denef:2008wq}. These provide an interpretation of gaugino condensation as arising from instantons, and each non-perturbative superpotential as a Pfaffian. 

We  define $\Im G\equiv\varphi$, which will be our EDE candidate. To realize EDE we fix the exponents $b,c,d,$ as:
\begin{equation}
c=2b \, , \qquad d=3b\,.
\end{equation}
The scalar potential for this model is given by,
\begin{equation}
V = V_{\rm KKLT} + V_{\rm EDE}\,,
\end{equation}
where we define,
\begin{equation}
V_{\rm KKLT} = \frac{a^2 A^2 e^{-2 a t}}{6 t}+\frac{a A^2 e^{-2 a t}}{2 t^2}+\frac{a A W_0 e^{-a t}}{2 t^2}+\frac{M^2}{12 t^2}\,,
\end{equation}
and, fixing $d=3b$ and $c=2b$,
\begin{eqnarray}
V_{\rm EDE} =&& - \frac{b^2 B^2 + c^2 C^2 + d^2 D^2}{24 \gamma t^2 }  \\
&&+ \frac{B}{2 t^2}\left( a A e^{- a t}  - \frac{1}{3 \gamma }b C - \frac{1}{B \gamma} b^2 C D\right) \cos (b \varphi) \nonumber \\
&& + \frac{C}{2 t^2}\left( a A e^{- a t} - \frac{1}{4}b^2 D\right) \cos (2 b \varphi) \nonumber \\
&& + \frac{D}{2 t^2} a A e^{- a t} \cos (3 b \varphi) \, .  \nonumber
\end{eqnarray}
Note that, in the previous expressions, we have assumed consistent truncation at $\Im T=0$ and $\Re G=0$.  The first term in $V_{\rm EDE}$ is clearly degenerate with the last term of $V_{\rm KKLT}$, and thus the uplift from AdS to dS can be realized by balancing the two terms. The remaining terms are degenerate with $W_0$ and $M$, and thus, provided that $B,C,D \ll W_0$, they can be considered to be a perturbative correction to the KKLT potential. These same terms can be used to realize the EDE potential, eq. \eqref{eq:VEDE}.

To ensure minimal backreaction on KKLT stabilization, and that the structure of $V_{\rm EDE}$ is insensitive to the value or possible cosmological evolution of $t$, we impose that the dominant terms in $V_{\rm EDE}$ are those whose scaling with $t$ is universal across the terms in $V_{\rm EDE}$. Each of these already appears in $V_{\rm KKLT}$. To this end, we consider the parameter regime
\begin{equation}\label{Pfaffians}
  B, C, D \ll A e^{- a t}\, .
\end{equation}
In this case, we may approximate the EDE potential as
\begin{equation}
V_{\rm EDE} \simeq \frac{a A e^{- a t}}{2 t^2} B \cos (b \varphi)+ \frac{a A e^{- a t}}{2 t^2} C \cos(2 b \varphi)  + \frac{a A e^{- a t}}{2 t^2} D \cos(3 b \varphi)\, ,
\end{equation}
which we may in turn write as 
\begin{equation}
V_{\rm EDE} \simeq V_0 \left[ {\cal B} \cos (b \varphi)+ {\cal C} \cos(2 b \varphi)  +  {\cal D} \cos(3 b \varphi)  \right]\, ,
\end{equation}
where we define $B = \epsilon\, {\cal B}$, $C=\epsilon \, {\cal C}$, and $D = \epsilon \, {\cal D}$, with ${\cal B,C,D}= {\cal O} (1)$, and define the EDE normalization
\begin{equation}
V_0 = \frac{a A e^{- a t}}{2 t^2} \epsilon \,.
\end{equation}
Matching the desired EDE normalization requires $V_0 \sim ({\rm eV})^4 \sim 10^{-108}$ in Planck units. Meanwhile, in the standard KKLT solution \cite{Kachru:2003aw}, one has  $t_*\sim 113$, $a=0.1$, and $A=1$, all in Planck units.
From this, one finds that the desired $V_0$ is realized for $\epsilon$ given by
\begin{equation}
\epsilon = 2.0 \times 10^{-98}\, ,
\end{equation}
or in other words,
\begin{equation}
\label{eq:BCD}
B, C, D \sim \left( 10^{- 24} {\rm eV} \right)^3\, .
\end{equation}
Meanwhile, $A e^{- a t} \approx 1.2 \times 10^{-5} M_{pl}^3 \approx  \left( 5.6 \times 10^{25} {\rm eV}\right )^3$. Thus the above is well within the regime of validity of our assumption that $B,C,D \ll A e^{- a t}$, providing an {\it a posteriori} justification for this simplifying assumption.  In this case, $t$ can be stabilized at $t_*$ as in KKLT, with negligible corrections due to the EDE.

Finally, to arrive at the EDE potential we first note the K\"{a}hler potential gives a rescaled kinetic term for $\varphi$ as
\begin{equation}
{\cal L}_{\varphi{\rm kin}} =  \frac{t_*}{3 (-\gamma)} (\partial \varphi)^2\, . 
\end{equation}
where we have fixed $t=t_*$, the position of the KKLT minimum. We therefore define the canonical EDE variable $\phi$ as
\begin{equation}
\phi \equiv  \sqrt{\frac{2\ t_*}{3 (- \gamma) }}\varphi \, ,
\end{equation}
and find the scalar potential for $\phi$ given by
\begin{equation}
V_{\rm EDE} = V_0 \left[ {\cal B} \cos (\phi/f) + {\cal C} \cos{(2\phi/f)} + {\cal D} \cos (3\phi/f) \right] \,,
\end{equation}
where the decay constant is
\begin{equation}\label{decayconstantST}
f \equiv \sqrt{\frac{3 (-\gamma) }{2\ t_*}} \frac{1}{b} \, .
\end{equation}
To engineer a Planckian decay constant, $f \sim M_{pl}$, we consider $b <1 $ such that $\sqrt{t_* }b \sim 1$. The parameters ${\cal B}, {\cal C}, {\cal D}$ are finally fixed by the EDE potential as 
\begin{equation}
{ \cal{B}} = - \frac{15}{4} \,,\qquad {{\cal C} }= \frac{3}{2} \,,\qquad {{\cal D}} = - \frac{1}{4}\,,
\end{equation}
leading to the desired EDE potential, eq.~\eqref{eq:VEDE}, upon fixing $M$ to set the cosmological constant.
 
  \begin{figure}[h!]
\centering
\begin{subfigure}{.5\textwidth}
    \centering
    \includegraphics[width=.9\textwidth]{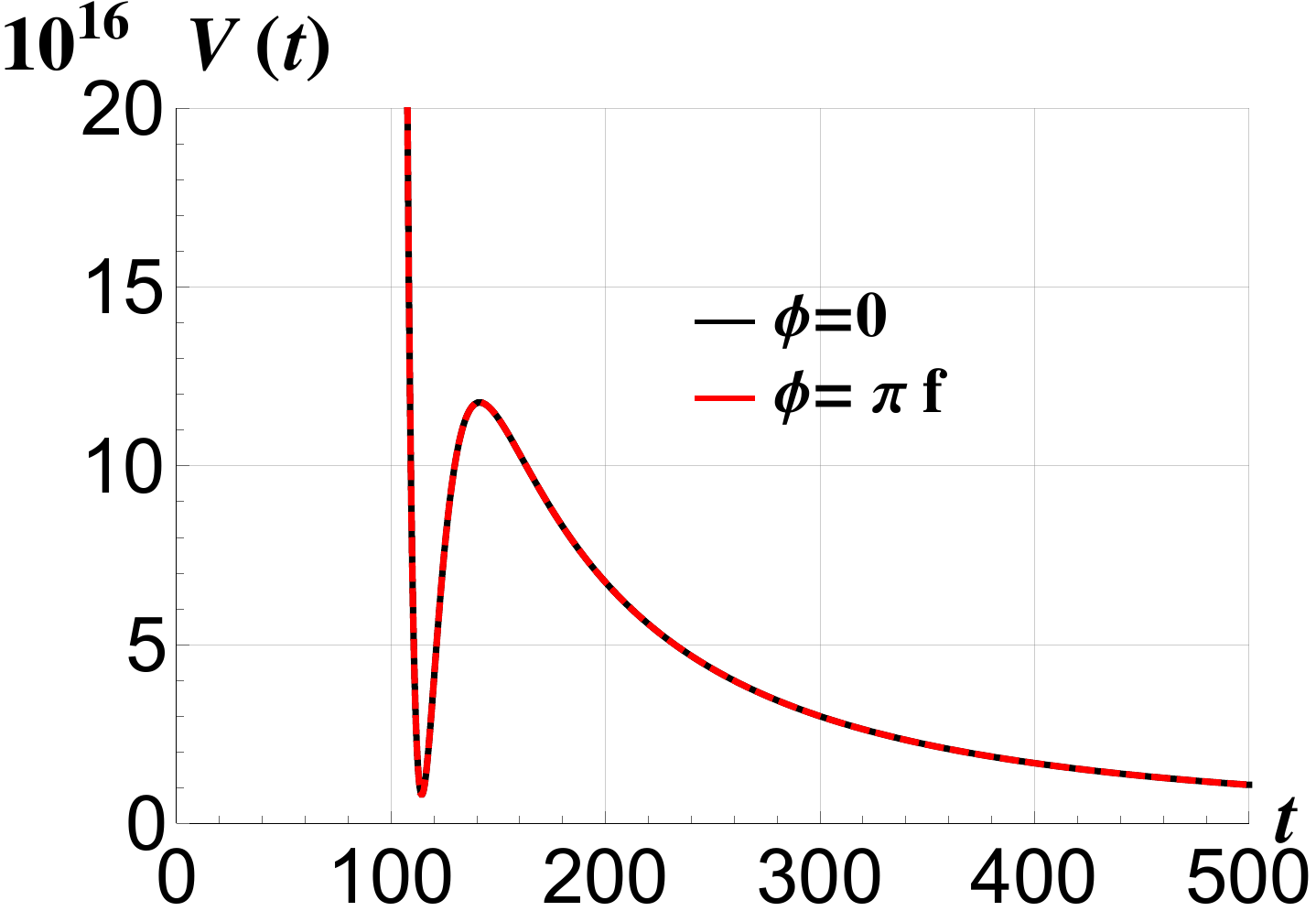}
    \caption{KKLT stabilization of the volume $t\equiv \Re T$.}
\end{subfigure}%
\begin{subfigure}{.5\textwidth}
    \centering
    \includegraphics[width=.9\textwidth]{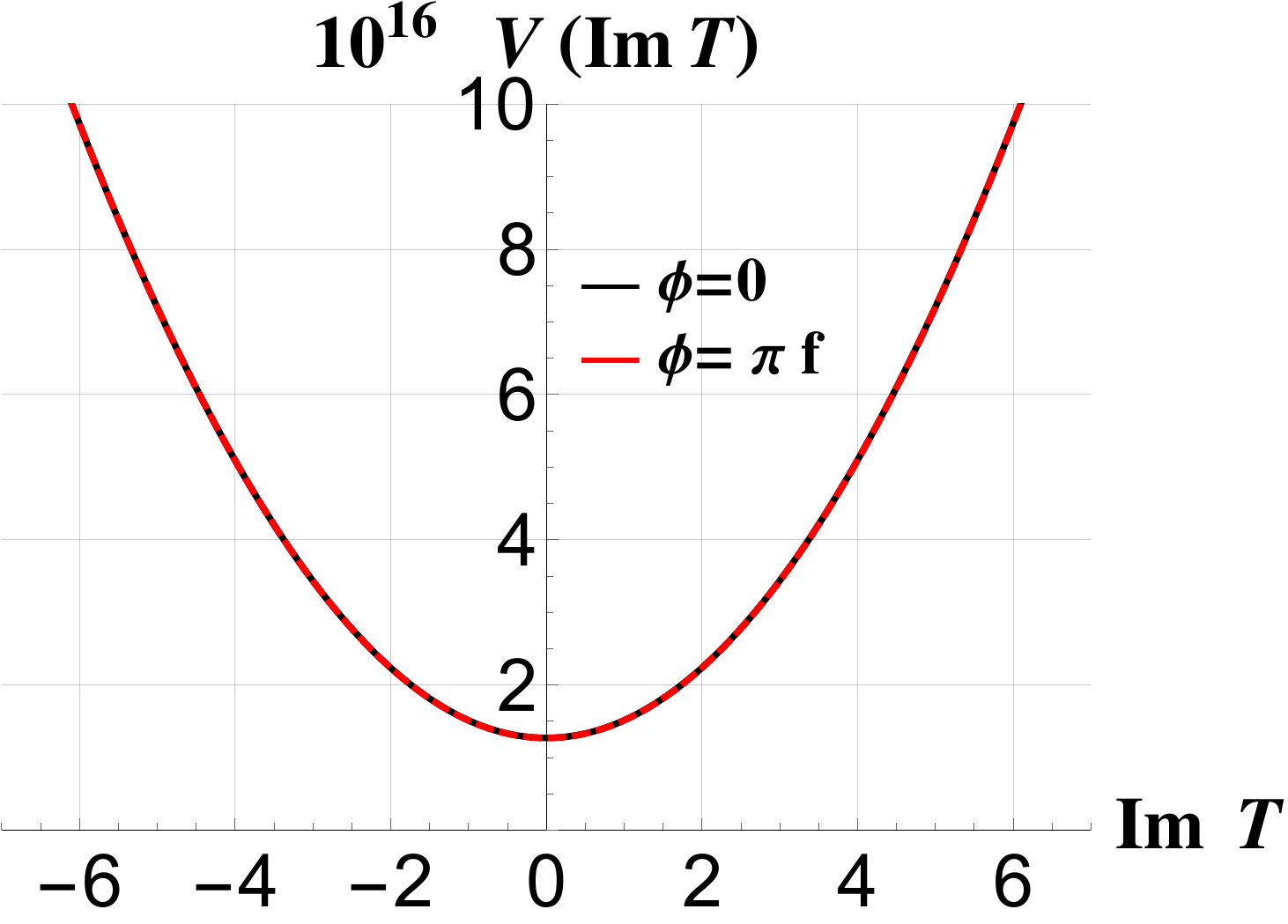}
    \caption{Stabilization of the $\Im T$.}
\end{subfigure} 
\par\bigskip 
\begin{subfigure}{.5\textwidth}
    \centering
    \includegraphics[width=.9\textwidth]{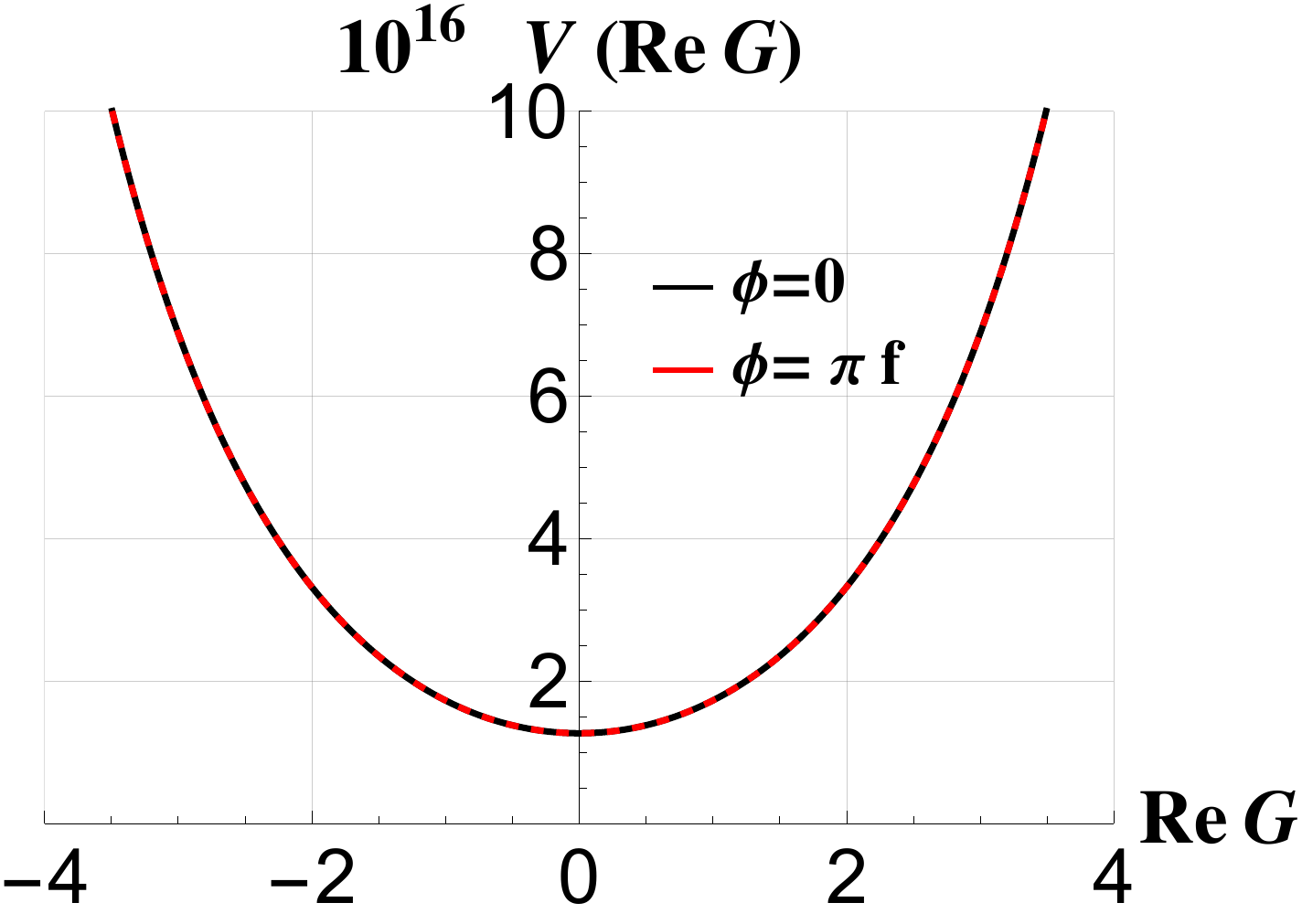}
    \caption{Stabilization of the $\Re G$.}
\end{subfigure}%
\begin{subfigure}{.5\textwidth}
    \centering
    \includegraphics[width=.9\textwidth]{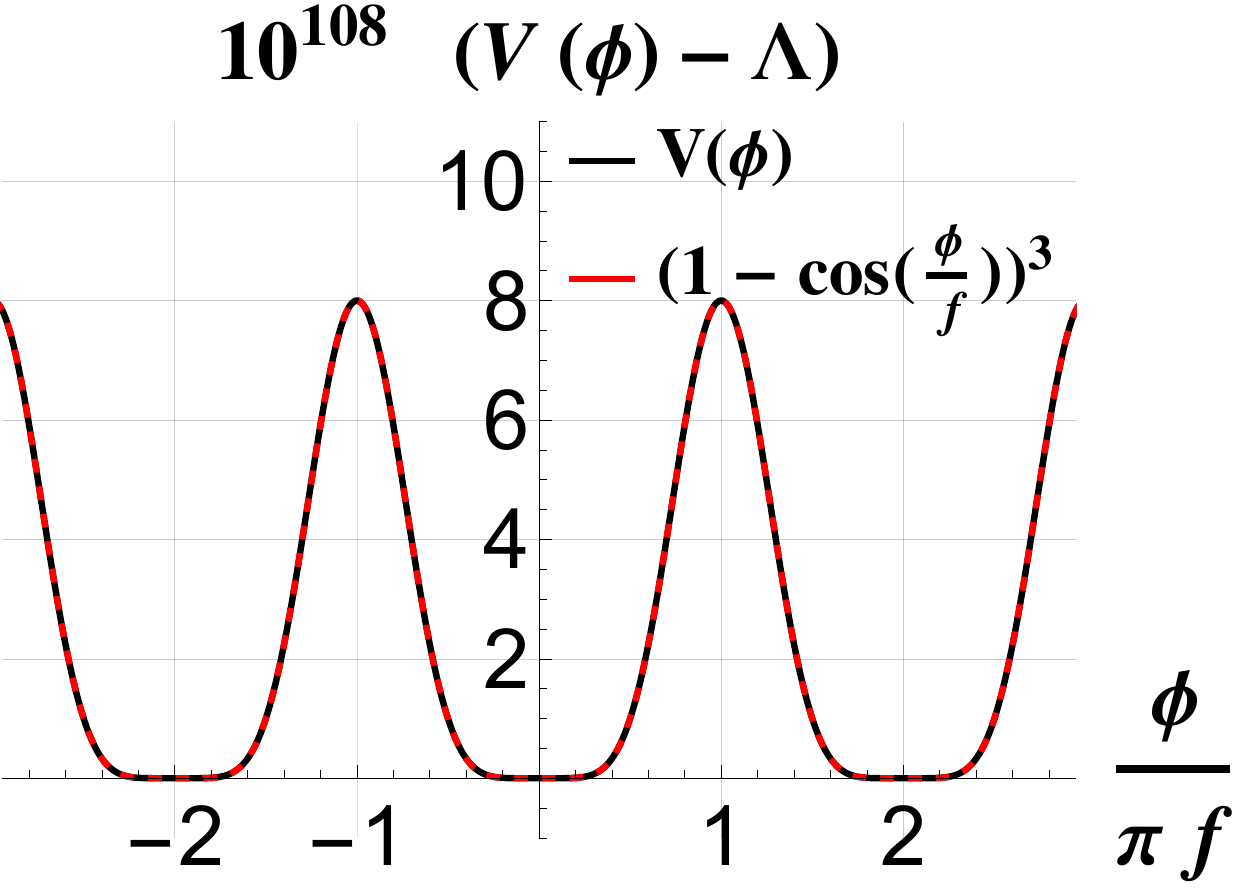}
        \caption{Early Dark Energy for $\phi$.}
\end{subfigure}
\par\bigskip 
\caption{  \label{fig:stabilization} Early Dark Energy and the Cosmological Constant in a fully-stabilized KKLT-type compactification. For $t\equiv \Re T$, $\Im T$, and $\Re G$, we show the potential when the EDE scalar $\phi$ is at the top (red, dashed) and bottom (black) of its potential. For the EDE scalar $\phi$, we compare to the desired $(1- \cos \phi/f)^3$ form. Parameters are given in \eqref{eq:BCD}, along with the standard KKLT parameters, $\gamma=-1$, and $b=1/\sqrt{t_*}$.  }
\end{figure}

We are now ready to explicitly demonstrate moduli stabilization and the realization of the EDE potential.  In Fig.~\ref{fig:stabilization}, we plot the potential for all four fields in our model: $\Re T\equiv t$, $\Im T$, $\Re G$, and the EDE field $\phi$ (canonically normalized  $\Im G$). From this one may conclude that the volume is stabilized as in KKLT, without any backreaction due to the EDE scalar. Subtracting off the cosmological constant and rescaling by $V_0 ^{4}$, one finds the desired oscillatory EDE potential for $\phi$. Meanwhile, $\Re G$ and $\Im T$ are stabilized at 0.  The gravitino mass $m_{3/2}\sim 6 \times 10^{10}$ GeV is set as in KKLT, where it is comparable to the barrier to decompactification \cite{Kallosh:2004yh} and generally `superheavy' by particle physics standards.

Finally, we would like to comment on some subtleties of this construction. First of all, the superpotential should receive contributions from Euclidean D1 (ED1) instantons on the two-cycle wrapped by the D5 branes. Following \cite{Ben-Dayan:2014lca,Cicoli:2021gss}, we have assumed these to be a subdominant effect, on the basis that ED1 corrections are suppressed at small string coupling and large volume \cite{Grimm:2007hs}. This issue will have to be addressed by a more explicit construction of the model. Second, our construction relies on the realization of the hierarchy eq.~\eqref{Pfaffians} for the Pfaffians of the non-perturbative terms. We remain agnostic about the precise mechanism through which such exponentially small parameters  could be engineered. An appealing possibility is via the competition of different  non-perturbative terms as studied in \cite{Demirtas:2021nlu,Demirtas:2021ote}.
 
 \section{Discussion}

The Early Dark Energy model fulfills its promise of uniting Planck CMB and SH0ES SNIa data, but it faces also a number of challenges, as outlined in the Introduction. It does not only require a fine-tuned potential but it also falls short of a concordance cosmological model \cite{Hill:2020osr,Ivanov:2020ril,DAmico:2020ods}, due to increased tensions with large-scale structure data. As a step towards interpreting and addressing these challenges, in this work we have presented realizations of Early Dark Energy in supergravity and also in a concrete string theory compactification setting.  In both cases, we have shown that the EDE scenario can co-exist with the realization of a small cosmological constant in a single consistent framework.

We have found that the $(1-\cos \phi/f)^3$ form of the EDE potential can be realized via a combination of multiple distinct non-perturbative effects, without requiring higher harmonics in an instanton expansion. In a simple supergravity toy model, we demonstrated that only two non-perturbative terms are needed, while in a KKLT-type compactification, preserving stabilization of the volume requires EDE be constructed by three distinct non-perturbative effects. 

This analysis has allowed us to reframe the challenges to EDE in terms of properties of the UV completion. The challenges outlined in the Introduction now take on a new form:
\begin{enumerate}

\item To correctly match the energy scale of the EDE one must be able to engineer small coefficients $B,C,D$, or equivalently the Pfaffian and hence energy scale of the gaugino condensation, to a very low scale. In order for this to survive beyond the probe approximation, SUSY breaking by the anti-D3 of KKLT will need to be hidden from the EDE brane stacks, so that SUSY localized on the brane is unbroken down to a low scale and we can use the conventional derivation of gaugino condensation. This `sequestering' of SUSY, while similar to that studied in detail in \cite{Burgess:2021juk,Burgess:2021obw}, is another model building challenge. The works \cite{Berg:2010ha,Berg:2012aq} might provide an interesting approach to possibly assessing this issue in the context of string compactification; another potential path to small $B,C,D$ is to work in analogy to \cite{Demirtas:2021nlu,Demirtas:2021ote}.
\item To get the form of the potential, in our simple string theory construction, we added to the standard KKLT setup an additional 3 brane stacks with $N_B$,  $N_C= N_B/2$, and $N_D = N_B/3 $ branes, where $N_B\equiv 2\pi/b = 66$ in our example. It remains a string model building challenge to construct a global compactification with such a large number of branes so delicately arranged, and with the D5 tadpole, eq. (4.20) of \cite{Grana:2005jc}, cancelled. We notice also that, for this example, the constraint on the decay constant $f<0.008$ of \cite{Rudelius:2022gyu}, corresponds to $b\approx14$, thus violating the minimal requirement $b\lesssim 2$ to obtain at least one brane for each stack.
\item Our string construction does not seem to immediately resolve the tension with LSS. Following \cite{Allali:2021azp,Ye:2021iwa},  additional ultra-light degrees of freedom might represent a solution to this problem. However, whereas the SUGRA toy model (Sec.~\ref{SEC:SUGRA}) provides naturally such a ultra-light particle as companion of the EDE scalar field in the complex multiplet,  we have shown that moduli stabilization (Sec.~\ref{SEC:STRING}) generically makes this saxion partner heavy. However, although this requires extra model-building, obtaining ultra-light particles in string compactification scenarios remains a very concrete possibility. An appealing option is to consider a $C_2$ axion along a different 2-cycle. In strongly warped throats, it has for example been argued \cite{Hebecker:2018yxs} that extremely light axions with masses scaling as the cube of the warp factor generically exists. They might represent ideal candidates to address this challenge.
\end{enumerate}

There are at least three distinct directions for future work: (1) string theory model building to engineer a concrete compactification with the desired properties, (2) cosmological data analyses of the models as they naturally arise in string theory, e.g., with three {\it a priori} unrelated non-perturbative terms without enforcing the cubed cosine structure on the potential, and (3) to explore the string theory embeddings of other model realizations of early dark energy, such as \cite{Alexander:2022own}. Finally, we note that recent work \cite{Kojima:2022fgo} provides a UV completion of EDE in a five-dimensional gauge theory, which may provide an alternate route to a string theory construction.  We leave these exciting directions to future work.
 
\acknowledgments 
 
The authors thank Heliudson de Oliveira Bernardo, J.~Colin Hill, James Halverson, Dragan Huterer, Liam McAllister, Michael Toomey, and Timm Wrase for insightful comments on a draft of this work. The authors thank Dragan Huterer for permission to reproduce Fig.~5  of \cite{HutererReview} for use as Fig.~\ref{fig:CMB-explainer} in this paper. EM is supported in part by a Discovery Grant from the National Sciences and Engineering Research Council of Canada.

\bibliography{refs}
\bibliographystyle{JHEP}

\end{document}